\documentclass[lettersize,journal]{IEEEtran}
\usepackage{amsmath,amsfonts,mathtools}
\usepackage{array}
\usepackage{textcomp}
\usepackage{stfloats}
\usepackage{url}
\usepackage{xurl}
\usepackage{verbatim}
\usepackage{graphicx}
\usepackage{cite}
\usepackage{xcolor}
\usepackage{booktabs}
\usepackage{tabularx}
\usepackage{multirow}
\usepackage{makecell}
\usepackage{xspace}
\usepackage{hyperref}
\usepackage{enumitem}
\usepackage{adjustbox}
\usepackage{pifont}
\usepackage[caption=false,font=normalsize,labelfont=sf,textfont=sf]{subfig}
\usepackage[frozencache,cachedir=minted-cache]{minted}
\usepackage{listings}
\usepackage[ruled,vlined,linesnumbered]{algorithm2e}
\usepackage[most]{tcolorbox}
\usepackage{cleveref}

\newcommand{\toolname}[0]{{\textsc{MASFuzzer}}\xspace}
\newcommand{\PromptFuzz}[0]{{\textsc{PromptFuzz}}\xspace}
\newcommand{\CKGFuzzer}[0]{{\textsc{CKGFuzzer}}\xspace}
\newcommand{\PromeFuzz}[0]{{\textsc{PromeFuzz}}\xspace}
\newcommand{\Hopper}[0]{{\textsc{Hopper}}\xspace}
\newcommand{\LibErator}[0]{{\textsc{LibErator}}\xspace}

\newcommand{\seqone}{API sequences from usage examples\xspace}
\newcommand{\seqTwo}{API Sequences via Mutation-Propagation-Grounded Mining\xspace}
\newcommand{\seqtwo}{API sequences via mutation-propagation-grounded mining\xspace}
\newcommand{\seqThree}{API Sequences via Semantic-Aware Mining\xspace}

\newcommand{\match}{API sequence mining\xspace}

\renewcommand{\footnoterule}{
  \kern -3pt
  \hrule width 0.4\columnwidth height 0.4pt
  \kern 2.6pt
}
\renewcommand{\arraystretch}{1.2}

\definecolor{light-gray}{gray}{0.87}
\newcommand{\graybox}[1]{\colorbox{light-gray}{#1}}

\begin{document}

\title{\Large \bf MASFuzzer: Fuzz Driver Generation and Adaptive Scheduling via Multidimensional API Sequences}



\author{
Xingyu Liu, Zengqin Huang, Xiang Gao, Hailong Sun
\thanks{Xingyu Liu, Xiang Gao, and Hailong Sun are with Beihang University, China (e-mail: lxingyu@buaa.edu.cn; xiang\_gao@buaa.edu.cn; sunhl@buaa.edu.cn).}
\thanks{Zengqin Huang is with Jinan University, China (e-mail: zqhuang@stu2022.jnu.edu.cn).}
\thanks{Corresponding author: Xiang Gao(e-mail: xiang\_gao@buaa.edu.cn).}
}



\maketitle

\begin{abstract}

Fuzz testing of software libraries relies on fuzz drivers to invoke library APIs. Traditionally, these drivers are written manually by developers---a process that is not only time-consuming but also often inadequate for exercising complex program behaviors. While recent studies have explored the use of Large Language Models (LLMs) to automate fuzz-driver generation, the resulting drivers still frequently fall short in covering deep program branches.
To address these challenges, we propose \toolname, a novel fuzzing framework that integrates multidimensional API sequence construction with adaptive fuzzing scheduling strategies to comprehensively enhance library testing.
At its core, \toolname synthesizes contextually relevant API call sequences by referring to the API usage examples from the codebase, and employing mutation-propagation-grounded and semantic-aware \match.

Such multidimensional API sequences serve as the foundation for LLMs to generate effective initial drivers.
In addition, \toolname incorporates a coverage-guided scheduler that prioritizes testing time for the most promising drivers, along with a driver-mutation strategy to evolve them, thereby systematically generating fuzz drivers to explore previously untested code regions.
We evaluated \toolname on 12 widely-used open-source libraries, and the results show that \toolname achieves 8.54\% more code coverage than state-of-the-art techniques. 
Moreover, \toolname successfully uncovered 16 previously unknown vulnerabilities in the extensively tested libraries (14 confirmed by the developer and 9 assigned with CVE IDs). These results demonstrate that \toolname provides a more efficient and practical solution for fuzzing software libraries.

\end{abstract}


\section{Introduction}

Fuzz testing (fuzzing) is one of the most effective techniques for identifying software vulnerabilities \cite{fioraldi2020aflpp, serebryany2016continuous, bohme2020boosting,manes2018art}. It works by providing a large volume of inputs to a target program, triggering potential abnormal behaviors \cite{chafjiri2024vulnerability}. Fuzzing libraries is of particular significance, as it enables systematic exploration of API behaviors and uncovers vulnerabilities that are otherwise difficult to detect through application-level testing \cite{ispoglou2020fuzzgen,liu2024afgen}. Since fuzz testing typically relies on a single program entry point, fuzzing a library requires an auxiliary program that instantiates objects and invokes APIs in a valid sequence. Such programs are commonly called \emph{fuzz drivers} \cite{serebryany2016continuous}. 

Traditionally, fuzz drivers are manually written by experienced developers. For example, Google’s OSS-Fuzz project \cite{serebryany2017oss} involves many developers to write fuzz drivers in order to integrate a wide range of open-source libraries. Nevertheless, this process is both time-consuming and labor-intensive. Moreover, the effectiveness of the resulting fuzz drivers heavily relies on the programming expertise of developers~\cite{liu2025llm4tdg,jeong2023utopia}. As a result, both academia and industry have devoted increasing efforts to automate fuzz-driver generation. 

Existing research on automated fuzz-driver generation follows two main paradigms: static analysis–based construction and LLM-based semantic synthesis. 
The first strand mines API usage sequences from concrete code examples (e.g., FUDGE \cite{babic2019fudge}) or model static constraint relationships among APIs (e.g., GraphFuzz \cite{green2022graphfuzz}) to construct fuzz drivers.
Although showing promising results, such methods inherently lack deeper semantic reasoning and are constrained by the limited diversity of existing usage patterns.
The second approaches automate fuzz-driver generation \cite{mallissery2023demystify}, relying on Large Language Models (LLMs), which have shown impressive capabilities in coding tasks. 
These methods range from directly prompting LLMs with API signatures (e.g., OSS-Fuzz-Gen \cite{Liu_OSS-Fuzz-Gen_Automated_Fuzz_2024} and PromptFuzz \cite{lyu2024prompt}) to more structured approaches that integrate external knowledge sources, such as API knowledge graphs (e.g., CKGFuzzer \cite{xu2025ckgfuzzer}) or retrieval-augmented generation from consumer code and documentation (e.g., PromeFuzz \cite{promefuzz-ccs25}). 

Despite these advances, existing approaches still face several fundamental limitations:
\begin{itemize}[leftmargin=*]
    \item \textbf{Limited and Fragmented API Context:} Existing approaches generate API sequences from a single source—static code patterns, syntactic constraints, or LLM-inferred semantics. This results in drivers that are either syntactically correct but semantically shallow, or semantically rich but unreliable. The absence of a unified, multi-faceted API context model limits the system's ability to generate drivers that are both syntactically diverse and semantically grounded.
    \item \textbf{Fixed and Uniform Execution Strategy:} Current systems treat all generated drivers equally, allocating the same fuzzing time and resources to each. They lack a mechanism to dynamically prioritize drivers that demonstrate higher coverage potential or to reallocate energy away from stagnant ones, leading to inefficient resource utilization.
\end{itemize}

To overcome the above limitations, we present \toolname, a novel framework for fuzz driver generation and scheduling.
First, to address the problem of a limited and static API context, we introduce a multidimensional API sequence construction method. Instead of relying on a single information source, we systematically synthesize API invocation contexts from three complementary perspectives: (1) extracting sequential patterns from real code examples to ensure realism, (2) leveraging type signatures to guarantee syntactic validity, and (3) employing LLM-based semantic analysis to propose novel, semantically plausible API combinations. This enriched, multidimensional context provides the foundation for generating drivers that are both robust and exploratory.

Second, to tackle the inefficient, uniform resource allocation for fuzzing each driver, we design a coverage-guided time scheduler. Unlike static scheduling, our scheduler continuously monitors each driver's runtime coverage feedback and dynamically prioritizes fuzzing energy. Drivers that demonstrate higher potential to uncover new code paths receive significantly more execution time, while stagnant drivers are deprioritized, ensuring optimal resource utilization throughout the campaign.
Moreover, we enhance driver evolution with a sequence-aware mutation strategy, which utilizes our multidimensional sequences to intelligently alter API call orders, enabling more focused exploration of uncovered program states.

We evaluated \toolname on 12 widely-used open-source libraries. In terms of code coverage, compared with state-of-the-art tools, \PromeFuzz~\cite{promefuzz-ccs25}, \CKGFuzzer~\cite{xu2025ckgfuzzer} and \LibErator~\cite{toffalini2025liberating}, \toolname achieves coverage improvements in 11 of these libraries, with an average increase of 8.54\%, 15.07\% and 37.95\%, respectively. We further conducted ablation studies to analyze the contributions of different components, including multidimensional API sequences, and the time scheduler and mutation strategy. The results demonstrate that each of these modules contributes to the overall effectiveness of \toolname to varying degrees. Although these libraries have been extensively tested, \toolname successfully discovered 16 previously unknown bugs in these libraries (14 confirmed by the developer and 9 assigned with CVE IDs).

In summary, we make the following contributions in this work:
\begin{itemize}[leftmargin=*, topsep=1pt]
\item[$\bullet$] We propose a novel approach for constructing multidimensional API sequences. By extracting usage examples from the original codebase and performing mutation-propagation-grounded and semantic-aware \match, we generate API sequences along three complementary dimensions, providing rich contextual information to guide LLMs in generating high-quality fuzz drivers.
\item[$\bullet$] We design and implement \toolname, a fuzzing framework that combines fuzz driver generation driven using multidimensional API sequences, a coverage-guided time scheduler and sequence-aware driver mutation, enabling more effective library testing.
\item[$\bullet$] We perform an extensive evaluation of \toolname across 12 widely-used libraries. The results demonstrate that our approach significantly improves code coverage compared to state-of-the-art techniques and has uncovered 16 previously unknown vulnerabilities (14 confirmed by the developer and 9 assigned with CVE IDs).
\end{itemize}

\section{Background and Motivation}

\subsection{Library Fuzzing} 

Fuzzing is a widely used software testing technique for automatically discovering vulnerabilities in programs~\cite{bohme2020fuzzing, zhao2024systematic}. It works by continuously generating large numbers of inputs and feeding them to the program under test (PUT) to expose crashes. Among various fuzzing paradigms, coverage-guided grey-box fuzzing (CGF) has become the de facto standard \cite{fioraldi2023dissecting,pham2019smart}. CGF uses lightweight instrumentation to collect code coverage feedback, which is then used to guide input generation and steer exploration towards deeper program behaviors \cite{wang2019superion}. 

While CGF is highly effective for standalone programs, fuzzing software libraries presents a unique challenge. Unlike programs with a clear entry point (e.g., a main function), libraries typically expose functionality only through their Application Programming Interface (API). As a result, fuzzing them requires a test harness, or fuzz driver, that specifies how to invoke the library functions. Manually writing such fuzz drivers is notoriously difficult and time-consuming. It demands a deep understanding of the library’s semantics, and then carefully handles the initialization and state management, and the construction of meaningful API call sequences. Any mistakes in fuzz drivers can easily limit the fuzzing effectiveness and even introduce spurious crashes.

Several approaches have been proposed for automated fuzz-driver generation. Some approaches rely on user-provided specifications of the APIs under test~\cite{libfuzzer, green2022graphfuzz}, while others learn API usage patterns from existing consumer code~\cite{babic2019fudge, ispoglou2020fuzzgen}. 
More recent works, such as \PromeFuzz~\cite{promefuzz-ccs25},  \CKGFuzzer~\cite{xu2025ckgfuzzer} and \PromptFuzz~\cite{lyu2024prompt}, leverage advances in Large Language Models (LLMs) to aid fuzz-driver generation. LLMs such as GPT-4 \cite{achiam2023gpt} and CodeLlama \cite{roziere2023code} have shown remarkable capabilities in understanding and generating human-like text and source code\cite{chen2021evaluating,cheng2025towards}. In addition, they can learn complex syntactic and semantic patterns from large code corpora, enabling applications in code generation and testing \cite{wang2023review, siddiq2024using, schafer2024empirical, liu2025can, xia2023keep,xia2024fuzz4all,meng2024large,hu2023augmenting}. As a result, LLMs have emerged as a promising alternative in this regard. \cite{eom2024fuzzing,deng2023large}



\subsection{Motivating Example}

Despite advances in LLM-based fuzz driver generation, constructing effective API sequences to explore deep program behaviors remains challenging. 
To motivate our approach, we present a buffer-overflow vulnerability in the \texttt{libplist} library, which was discovered by \toolname but missed by the previous techniques, including \PromptFuzz~\cite{lyu2024prompt}, \CKGFuzzer~\cite{xu2025ckgfuzzer}, and \PromeFuzz~\cite{promefuzz-ccs25}.
The corresponding fuzz driver generated by \toolname is shown in \autoref{fig:motivation-example}.

To expose this vulnerability, the fuzz driver first parses \verb|XML| input (i.e., \textit{data}) and converts it into \verb|plist| structure (lines 3--4), copies it (lines 5--6), and converts the copy into an \textit{openstep} representation (lines 7--10). 
The issue occurs during the copy at line 6: if the \verb|XML| input contains \texttt{\textbackslash 0} bytes, \textit{plist\_copy} allocates memory and copies data only up to the null byte, since it relies on \texttt{strdup} for allocation. 
However, the internal length metadata of the \verb|plist| node preserved during the copy still records the original XML length, rather than the truncated size. As a result, when \texttt{plist\_to\_openstep} is invoked at line 10, it steers its data access based on its internal metadata, thereby causing it to read beyond the allocated buffer and triggering the vulnerability.

\begin{figure}[t]
\begin{minted}[
    fontsize=\footnotesize,
    breaklines,
    linenos,
    numbersep=2pt,
    xleftmargin=5pt,
    escapeinside=||
]{c}
int LLVMFuzzerTestOneInput(uint8_t *data, size_t size) {
  plist_t xml_plist = nullptr;
  if (|\graybox{plist\_from\_xml}|((const char*)data, (uint32_t)size, &xml_plist) == PLIST_ERR_SUCCESS && xml_plist) {
    plist_type node_type = plist_get_node_type(xml_plist);
    if (node_type != PLIST_NONE) {
      plist_t copied_plist = |\graybox{plist\_copy}|(xml_plist);
      if (copied_plist) {
        char* openstep = nullptr;
        uint32_t openstep_len = 0;
        if (|\graybox{plist\_to\_openstep}|(copied_plist, &openstep, &openstep_len, 0) == PLIST_ERR_SUCCESS) {
          if (openstep) free(openstep);
        }
        plist_free(copied_plist);
      }
    }
    plist_free(xml_plist);
  }
  return 0;
}
\end{minted}
\caption{Simplified fuzz driver generated by \toolname that exposes a previously unknown buffer overflow in \texttt{libplist}.}
\label{fig:motivation-example}
\end{figure}

Based on this fuzz driver, the invoked API sequence exhibits two critical categories of dependencies. The first is the \textit{type dependency}, where the return type of \texttt{plist\_copy} directly matches the parameter type of \texttt{plist\_to\_openstep}, ensuring that data can correctly flow across APIs to form a valid data flow. Moreover, there is the \textit{semantic dependency}, where \texttt{plist\_from\_xml} must be invoked before \texttt{plist\_copy} so that the \verb|XML| input is properly parsed into a \verb|plist| that is the required argument for \texttt{plist\_copy}. This semantic dependency is not only enforced by type constraints but also arises from the correct ordering of API usage. 

However, existing techniques fail to accurately infer and cover these dependencies in the fuzz drivers they generate. 
\PromptFuzz only provides selected library API functions and then prompts the LLMs to generate fuzz drivers. Since there is no detailed guidance, LLMs struggle to capture all the type dependencies and enforce correct API ordering. \CKGFuzzer attempts to mitigate this issue by constructing function-level call graphs, while \PromeFuzz computes different relevance scores among APIs and selects highly correlated APIs to compose fuzz drivers. Although these strategies can partially mitigate the problem, they still miss many sequential dependencies. 
For the example in \autoref{fig:motivation-example}, reaching the bug requires the sequence from \texttt{plist\_from\_xml} to \texttt{plist\_copy}, but there is no direct function call relationship between them. Consequently, \CKGFuzzer, which relies on call graphs, fails to capture this connection. Similarly, \PromeFuzz overlooks this dependency; its relevance-based selection filters out the generic \texttt{plist\_copy} due to its weak association with the core conversion task.

\begin{figure*}[t]
    \centering
    \includegraphics[
        page=1,
        trim=0in 1in 0in 0in,
        clip,
        width=\textwidth
    ]{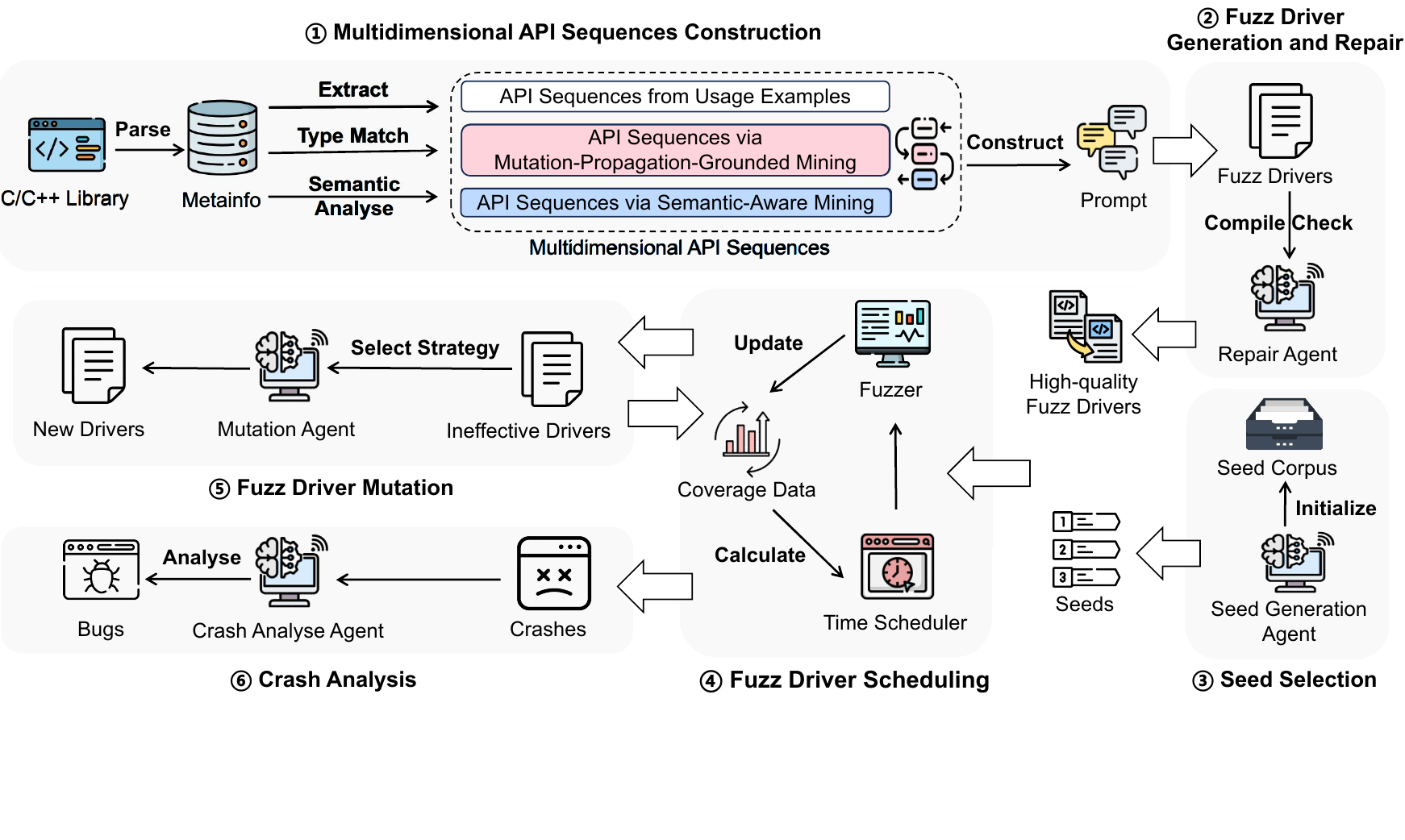}
    \caption{Overall workflow of \toolname.}
    \label{fig:workflow}
\end{figure*}

Motivated by this insight, we propose \toolname, which incorporates multidimensional API sequences during fuzz-driver generation. In the example shown in \autoref{fig:motivation-example}, \toolname is able to capture type dependency between \texttt{plist\_copy} and \texttt{plist\_to\_openstep}, forming one dimension of the sequence, as well as the semantic dependency between \texttt{plist\_from\_xml} and \texttt{plist\_copy}, forming another dimension. By providing the above two dimensions of correlated API sequences in the LLM's context, \toolname enhances semantic richness while ensuring syntactic correctness. This enables it to generate effective fuzz drivers and, by allocating more execution time to this driver through the dynamic scheduling strategy, successfully produce XML inputs containing null bytes, thereby covering previously unreachable code branches and ultimately discovering previously unknown bugs.

\section{Methodology}

\subsection{Overview}

\toolname is a general-purpose fuzzing framework designed to automatically generate high-quality fuzz drivers, and to perform dynamic scheduling and mutation during execution. Its overall workflow is illustrated in \autoref{fig:workflow}, which consists of six stages. At stage \ding{172}, \toolname first constructs API sequences---temporal chains of API calls---to guide fuzz driver generation. These sequences are produced in three complementary ways: (1) extracting real-world usage patterns for reliability, (2) composing type-compatible APIs, and (3) leveraging LLMs for semantic analysis to explore new logical flows.
This multidimensional approach captures both real-world usage patterns and potential novel scenarios.

These generated API sequences with type information and semantic context are used to construct LLM prompts and generate candidate fuzz drivers. To ensure correctness, at stage \ding{173}, a dedicated verification agent automatically repairs compilation errors by iteratively analyzing failed builds and applying corrective patches. For each generated fuzz driver, \toolname constructs an initial seed corpus at stage \ding{174} by extracting information from documentation.

After that, with the generated fuzz drivers and the selected seeds, \toolname implements a runtime driver scheduling mechanism to dynamically assign execution time to each driver at stage \ding{175}. The scheduler evaluates fuzz drivers based on their sequence-level similarity and API-level coverage. The scheduler prioritizes drivers with lower coverage while skipping those that have already reached a predefined threshold to avoid redundant testing. After this stage, the fuzz drivers are executed by the fuzzer. 
If a driver fails to uncover new branches (stage \ding{176}), \toolname evaluates the fuzzing potential energy and coverage of their associated APIs, and selects a mutation strategy that emulates the seed mutation process in fuzzing to generate a new fuzz driver. In addition, \toolname introduces a crash analysis agent that leverages LLM to systematically analyze fuzz drivers and crash information, distinguishing between API misuse and real library bugs (stage \ding{177}).

In the following, we mainly introduce how \toolname generates API sequences (stage \ding{172}), schedules different fuzz drivers (stage \ding{175}), and mutates them (stage \ding{176}) in more detail. For the other stages \ding{173} \ding{174} and \ding{177}, \toolname adopts similar strategies to those used in existing works~\cite{xu2025ckgfuzzer, lyu2024prompt, promefuzz-ccs25}.

\subsection{Multidimensional API Sequences}
\label{sec:mas}

Multidimensional API sequences model valid API usage orders, providing the domain-specific insights needed for fuzz driver generation. The process begins with a syntactic analysis of the target library to extract function-level \textit{metainfo}---including signatures, bodies, file names, line numbers, and documentation comments. This metadata forms the basis for constructing valid API sequences from three distinct and complementary perspectives.


\subsubsection{\textbf{API Sequences from Usage Examples}}  Within the library source code itself, sequences of public APIs are frequently invoked. 
These invocations---commonly found in example programs, unit tests, and fuzzing tests---implicitly reveal the correct usage patterns and constraints of public APIs.
\autoref{fig:libxlsxwriter_example} shows a typical usage scenario implemented by developers in the \verb|libxlsxwriter| library, which 
is a widely used library for programmatically generating Excel (XLSX) files. 
This example illustrates a representative API usage sequence, covering object creation, format configuration, data and formula writing, and finally closing the workbook to release resources and generate the file.

\begin{figure}[h]
    \centering
    \begin{minted}[
        fontsize=\small,
        breaklines,
        linenos,
        numbersep=4pt,
        xleftmargin=8pt,
        escapeinside=||
    ]{c}
int main() {
    lxw_workbook *wb = |\textbf{workbook\_new}|(FILE);
    lxw_worksheet *ws = |\textbf{workbook\_add\_worksheet}|(wb, NULL);
    lxw_format *fmt = |\textbf{workbook\_add\_format}|(wb);
    |\textbf{format\_set\_bold}|(fmt);
    |\textbf{worksheet\_write\_number}|(ws, 0, 1, 0, NULL);
    ...
    |\textbf{worksheet\_write\_array\_formula}|(ws, RANGE("A1:A3"), "{=SUM(B1:C1*B2:C2)}", fmt);
    return |\textbf{workbook\_close}|(workbook);
}
    \end{minted}
    \caption{API sequence extracted from usage example.}
    \label{fig:libxlsxwriter_example}
\end{figure}

Formally, we define the \textit{API sequences from usage examples} as follows. Given the code repository of a library, let the set of public APIs be \(\mathcal{A} = \{ a_1, a_2, \ldots, a_m \}\). 
\toolname first performs the abstract-syntax-tree (AST) and control-flow-graph (CFG) parsing to extract sequences of APIs invoked in the source code of the library itself. 
Within in a sequence of invoked APIs $S$, if there exist at least two invoked APIs \( a_i, a_j \in S \) such that \( a_i, a_j \in \mathcal{A} \), then we consider that there exists a useful API sequence within \( S \), represented as an ordered set \( Seq_{ue} \), where $Seq_{ue}$ is a subsequence of $S$ and \( \{\forall a_i \in Seq_{ue}\ |\ a_i \in \mathcal{A}\} \).
The order of elements in \( Seq_{ue} \) reflects the sequential invocation order of the APIs from real developers.
These sequences are directly extracted from authentic source code and are explicitly constructed and validated by developers, thereby embodying the most semantically rich and practically reliable API usage patterns.
However, since these sequences are derived from existing usage patterns, they are primarily valid sequences, but exhibit limited capability in discovering new coverage. Unlike FUDGE~\cite{babic2019fudge} and FuzzGen~\cite{ispoglou2020fuzzgen}, which directly splice extracted sequences to generate code, \toolname treats these API sequences merely as an exemplary skeletal reference. By deeply fusing them with multidimensional API sequences, \toolname leverages LLMs to effectively break the limitations of static real-world use cases.

\subsubsection{\textbf{\seqTwo}}  A well-constructed fuzz driver must possess a key property that we call Mutation Propagation Completeness (MPC). MPC ensures that:
(1) The fuzzer input can thoroughly exercise the entire API sequence; (2) each API call is sensitive to input variations, i.e., distinct inputs alter the API call's output or internal state; and (3) these changes propagate through the sequence, enabling that a single mutation can influence subsequent behaviors and reach deeper program states.

To satisfy the MPC property, we propose a \seqtwo paradigm. Specifically, when the return type of an API \textit{A} is type-compatible with the parameter type of another API \textit{B}, it indicates that the output of \textit{A} can be propagated to \textit{B}, thereby forming an inter-API data transfer channel.  Formally, we provide the definition and construction process of \seqtwo:

\textbf{(1) Type Compatibility Definition.} For two API \( a_i \) and \( a_j \), compatible(\allowbreak$a_i$\allowbreak,\allowbreak$a_j$\allowbreak) is true if and only if:
\[
\exists \, p_j \in \text{Params}(a_j) \quad \text{such that} \quad \text{Type}(\text{Return}(a_i)) = \text{Type}(p_j)
\]
Here, $\text{Type}(p_j)$ denotes the type of parameter $p_j$, indicating that the output of $a_i$ can flow into the input of $a_j$.

\textbf{(2) Compatibility Graph Construction.} Given an API set \(\mathcal{A} = \{a_1, a_2, \dots, a_n\}\), their compatibility graph a directed graph \( G = (V, E) \):
\[
G=\begin{cases}
V = \mathcal{A} \\
E = \{(a_i, a_j) \mid \texttt{compatible}( a_i, a_j )\ \text {for}\ a_i, a_j \in \mathcal{A}\}
\end{cases}
\]
For input parameters and return values that are difficult to parse due to complex macro definitions, we leverage LLMs to perform more complete and reliable supplementary analysis.

Note that, we primarily focus on the type-compatible relationships between structs and classes. For primitive data types (e.g., char and int), we assume that these values can be obtained directly or indirectly through input data mutation during fuzzing.
Moreover, including primitive types in type-compatibility analysis could introduce excessive false-positive propagation edges, thereby diluting the precision of our sequence mining.

\textbf{(3) Sequence Generation.} 
We traverse the compatibility graph \( G \) using a depth-first search (DFS). The traversal begins with nodes with zero in-degree and constrains a maximum sequence length of \( L \). The resulting set of API sequences is defined as:
\[
 Seq_{mp} = \bigcup_{k=1}^{m} \text{DFS}(v_k, \, G, \, L), \quad \forall \, v_k \in \{v \mid \text{indegree}(v) = 0\}
\]

\(\text{DFS}(v_k, G, L)\) denotes a depth-first search starting from node \(v_k\), which terminates either when a leaf node is reached or when the path length reaches the threshold \(L\). To effectively mitigate potential path explosion, we enforce strict cycle avoidance during traversal to prune redundant loops. Subsequently, we randomly sample the extracted valid paths to form the final set \( Seq_{mp} \), which represents a diverse pool of viable API sequences that satisfy the type-compatible propagation constraint.

\subsubsection{\textbf{\seqThree}} 

Overall, API sequences constructed via mutation-propagation analysis enable the synthesis of API sequences that are not explicitly present in the original codebase but may still be functionally valid. 
However, this approach relies heavily on the accurate parsing of parameter and return types. 
Consequently, APIs that lack input arguments, return \texttt{void}, or whose input and output types do not satisfy type-compatibility criteria cannot be incorporated into API sequences.
Moreover, sequences formed solely by type analysis may not always be semantically meaningful.
A sequence might combine APIs in a syntactically valid but logically incoherent manner---such as creating an object and immediately deleting it, without any meaningful operations in between.
To solve this problem, we further propose API sequence generation from semantic analysis.

To systematically derive semantically coherent API sequences, the construction workflow is structured into three stages: \textit{semantic extraction}, \textit{semantic relationship identification}, and \textit{sequence synthesis}.
First, in the semantic extraction stage, LLM is prompted with API metadata (e.g., signatures and documentation snippets) and code to obtain structured semantic descriptions for each API.
These descriptions summarize the functional role and usage context of the API, serving as semantic anchors for subsequent analysis.
Second, during relationship identification, \toolname feeds both the original API metadata and the extracted semantic descriptions into a relationship-prediction prompt.
This prompt guides the LLM to infer semantically meaningful predecessor-successor relationships between APIs---capturing dependencies, logical ordering, and contextual relevance that are often missed by purely syntactic or type-based methods.
Finally, \toolname utilizes the identified semantic relationships to construct API sequences \( Seq_{sem} \). 

\begin{figure}[h]
    \centering
    \begin{minted}[
        fontsize=\small,
        breaklines,
        numbersep=2pt,
        xleftmargin=5pt,
        escapeinside=||  % 添加这一行以启用 LaTeX 命令插入
    ]{c}

  |\textbf{ares\_library\_init}|(...)|$\rightarrow$|  |\textbf{ares\_init}|(...)|$\rightarrow$|
  |\textbf{ares\_gethostbyname}|(...)|$\rightarrow$|  |\textbf{ares\_destroy}|(...)|$\rightarrow$|
  |\textbf{ares\_library\_cleanup}|(...) 
    \end{minted}
    \caption{API sequence from semantic analysis.}
    \label{fig:c-ares_example}
\end{figure}

\autoref{fig:c-ares_example} demonstrates an API sequence via semantic-aware mining. All APIs in this sequence originate from the \texttt{c-ares} library, which is an asynchronous DNS resolver. In this example, the correct execution of the function \texttt{ares\_gethostbyname} depends on the prior invocation of the following two functions: \texttt{ares\_library\_init}, which correctly initializes the global library state, and \texttt{ares\_init}, which properly initializes the channel variable.

Notably, this dependency relationship is not conveyed through explicit parameters or return values, making it unrecognizable by conventional type analysis.
Instead, their effects manifest implicitly via global state changes and variable initialization. As a result, static analysis fails to establish the calling relationship among these APIs. 
In contrast, semantic analysis leverages an understanding of APIs' functional descriptions and semantic relationships to accurately identify that \texttt{ares\_gethostbyname} semantically depends on prior global initialization. This enables constructing the correct invocation order and dependency path.

\subsubsection {\textbf{Integrated API Sequences for Fuzz Driver Generation}} 

Overall, the three sequence types serve complementary roles in fuzz driver generation. Usage-example sequences \( Seq_{ue}\) capture common and reliable invocation patterns observed in real-world code, providing a stable backbone for driver construction.
Mutation-propagation sequences \( Seq_{mp}\) extend beyond typical usage by modeling inter-API data-flow compatibilities, exposing previously unexplored yet syntactically feasible call chains with high coverage potential. 
Semantic sequences \( Seq_{sem}\) incorporate contextual dependencies and logical constraints that static analysis cannot capture, thus mitigating API isolation and enriching the sequence space with semantic-aware variations.

To synthesize high-quality fuzz drivers, we integrate these multidimensional sequences into a structured prompting strategy. 
For each target API, we select one representative, previously unused sequence of maximal length from each of the three categories. Each sequence type is assigned a distinct role in the prompt:
\( Seq_{ue}\)  serves as the structural backbone, preserving realistic and stable invocation patterns;
\( Seq_{mp}\)  acts as a coverage-extending variant, introducing type-compatible call reorderings that break out of common usage boundaries;
\( Seq_{sem}\) functions as a semantic complement, injecting context-sensitive dependencies that guide the LLM toward semantically consistent API compositions.
For a certain API, if the API sequence of any category is missing, \toolname just relies on the sequences from available categories.

A prompt instructs the LLM to combine these three perspectives---preserving the natural ordering from \( Seq_{ue}\), incorporating mutation-aware extensions from \( Seq_{mp}\), and respecting semantic constraints from \( Seq_{sem}\)---to generate a unified fuzz driver that is both syntactically correct and semantically directed toward deeper state exploration.

\subsection{Coverage-Guided Time Scheduler}

Most fuzz driver frameworks~\cite{xu2025ckgfuzzer, lyu2024prompt, chen2023hopper, promefuzz-ccs25} assume that all generated drivers are equally effective in testing and, therefore, allocate an identical execution time to each driver. However, in practice, different fuzz drivers exhibit substantially different testing potentials. Ideally, drivers with higher potential should be assigned more execution time to explore deeper program states and exercise hard-to-reach branches. Furthermore, code coverage generally follows the law of diminishing marginal returns~\cite{zhu2022fuzzing}.
As fuzzing progresses, APIs targeted by later drivers are often already extensively covered during earlier testing phases. Under such circumstances, continuing to allocate time uniformly across all drivers exacerbates resource inefficiency and leads to a fundamental dilemma: on the one hand, high-potential drivers cannot obtain additional execution time to fully show their exploration capability; on the other hand, drivers scheduled later may be constrained by the remaining time budget, preventing them from triggering complex and deep execution paths.

\begin{algorithm}[h]
\caption{Coverage-Guided Time Scheduling}
\label{alg:time-scheduling-algorithm}

$\mathit{Cov} \leftarrow \text{initCov}()$\;
$\mathcal{S}_{exec} \leftarrow \emptyset$ \;
\ForEach{$i \in \text{range}(0, |D|)$}{
    $\mathit{uniqueAPIs} \leftarrow \text{removeDuplicates}(D[i])$\;
    $\mathit{avgCov} \leftarrow \frac{\sum_{\mathit{api} \in \mathit{uniqueAPIs}}\text{\textbf{covered}}(\mathit{Cov}[\mathit{api}])}{\sum_{\mathit{api} \in \mathit{uniqueAPIs}}\text{\textbf{total}}(\mathit{Cov}[\mathit{api}])}$\;

    \If{$\mathit{avgCov} > \theta$}{
        \textbf{continue}\;
    }
    \Else{
        $\alpha_t \leftarrow \text{Bound} ( B^{i/|D| - 1},\; \alpha_{\min},\; \alpha_{\max} ) $\;
        $t \leftarrow T / |D| \times \alpha_t$\;
        $\omega_{novelty} \leftarrow \min\limits_{\mathclap{S \in \mathcal{S}_{exec}}} \text{Lev}_{w}(\mathit{Seq}(D[i]), S) / |\mathit{Seq}(D[i])| $\;
        $t_{assign} \leftarrow (t / \max(\mathit{avgCov}, \beta)) \times \omega_{novelty}$\;
    }

    $t_\mathit{actual}, \mathit{newCov} \leftarrow \text{ExecuteFuzzDriver}(D[i], t_{assign})$\;
    
    \If{$\mathit{newCov} = \emptyset$}{
        $D'[i] \leftarrow \text{Mutate}(D[i])$ \tcp{\autoref{mutation}}
        $t_\mathit{actual}, \mathit{newCov} \leftarrow \text{ExecuteFuzzDriver}(D'[i], t_{assign})$\;
    }
    
    $T \leftarrow T - t_\mathit{actual}$\;
    $\mathit{Cov} \leftarrow \mathit{Cov} \cup \mathit{newCov}$\;
    $\mathcal{S}_{exec} \leftarrow \mathcal{S}_{exec} \cup \{ \mathit{Seq}(D[i]) \}$\;
}
\end{algorithm}

To address these challenges, we propose a coverage-guided time scheduler that dynamically adjusts the execution time of each fuzz driver. Before a driver is executed, the scheduler evaluates its potential testing value based on current multi-dimensional sequence-level and API-level coverage feedback, and allocates an appropriate runtime budget accordingly. This design ensures that each driver is assigned an execution time proportional to its remaining fuzzing potential. The details of the scheduling algorithm are illustrated in \autoref{alg:time-scheduling-algorithm}.

The coverage-guided time scheduling algorithm processes each fuzz driver through a systematic procedure. For the $i$-th driver $D[i]$, the algorithm first extracts unique APIs through deduplication.
It then calculates the current average coverage $\mathit{avgCov}$, where $\mathit{Cov}$ provides historical coverage data for each API.
Drivers exceeding the coverage threshold $\theta$ are immediately skipped to optimize resource utilization. For drivers requiring execution, the algorithm computes a time coefficient $\alpha_t$ using an exponential function based on the proportion of remaining drivers ($i/|D| - 1$), bounded between $[\alpha_{\min}, \alpha_{\max}]$. This approach, inspired by the simulated annealing algorithm used by AFLGo~\cite{bohme2020boosting}, ensures balanced time allocation throughout the fuzzing process.

The scheduler introduces a novelty factor $\omega_{novelty}$, to effectively leverage sequence information for further capturing the testing potential of fuzz drivers. This factor is computed based on the minimum weighted edit distance between the current sequence and the set of historical sequences, where dimension-specific weights are applied and normalized across different sequence dimensions. Fuzz drivers whose execution paths are highly similar to previously explored ones are regarded as redundant and thus allocated less execution time.

The base time allocated $t$ is derived from the proportional time share adjusted by $\alpha_t$. The final assigned time $t_{\text{assign}}$ is then computed by inversely scaling with the coverage.
Higher coverage results in proportionally less execution time, with a lower bound determined by $\beta$ to prevent excessive time allocated to low-coverage drivers.

After time assignment, the driver executes for the specified duration, returning actual execution time $t_{\text{actual}}$ and new coverage data $\mathit{newCov}$. 
The driver that fails to discover new branches then undergoes mutation using predefined strategies and is re-executed, followed by updates to the remaining total time $T$ and the aggregated coverage information $\mathit{Cov}$.
This strategy is designed to achieve an optimal balance between thorough testing and time efficiency by prioritizing drivers with lower coverage while maintaining a fair resource distribution across all drivers.

\subsection{Sequence-Aware Mutation Strategy}
\label{mutation}

To enhance fuzzing coverage, \toolname automates the generation of diverse fuzz drivers, aiming to explore more execution paths. However, not all of the generated drivers are effective---some fail to reach new execution paths or trigger potential crashes during fuzzing. To address these ineffective drivers, we propose a novel fuzz driver iterative refinement mechanism that integrates coverage-guided mutation with multidimensional API sequence features, enabling the iterative generation of more effective and diverse fuzz drivers. The mutation strategy consists of two stages: energy allocation scheme and mutation strategy selection. 

\textbf{Energy Allocation Scheme.} Previous fuzzing frameworks~\cite{xu2025ckgfuzzer, lyu2024prompt, bohme2016coverage} have introduced the concept of \textit{energy} to represent the execution potential of an API, where APIs with higher energy are prioritized during fuzz driver construction. 
However, traditional energy calculations often rely solely on coverage, while overlooking how an API is used across potential API sequences. 
To address this limitation, we introduce a new metric called \textit{potential}, which characterizes the usage potential of an API based on the number of distinct API sequences in which it appears.
Formally, the novel energy computation formula is defined as follows:
\[
\text{energy(\textit{api})} = (1 - \text{cov(\textit{api})}) \times (1 - \text{freq(\textit{api})}) \times \text{potential(\textit{api})}
\]
Here, \textit{cov(api)} denotes the branch coverage ratio of the API, reflecting the proportion of its branches that have already been exercised—higher values indicate better-tested APIs, and \textit{freq(api)} represents the proportion of executed fuzz drivers that invoke this API—higher values imply more frequent usage.
Moreover, \textit{potential(api)} captures the number of multidimensional API sequences in which the API appears, reflecting its semantic diversity and usage richness. Together, these three components integrate both static and dynamic aspects of API behavior to quantify its energy for guiding fuzz driver refinement.

\textbf{Mutation Strategy Selection.} Traditional mutation strategies in fuzzing primarily target the input data fed into the program under fuzz. In contrast, \PromptFuzz extends the mutation process to the level of API within the fuzz driver itself. Building upon this insight, we propose an enhanced sequence-level \textit{Mutation Strategy Selection} mechanism that incorporates multidimensional API sequences (\( Seq_{ue} \cup Seq_{mp} \cup Seq_{sem} \)) to guide the generation of more effective fuzz drivers.
Specifically, during each mutation iteration, our approach first identifies the API with the highest energy and retrieves an unused API sequence associated with it, and then randomly selects one of three mutation strategies—\textit{Insert}, \textit{Replace}, or \textit{Combine}.
\begin{itemize}[leftmargin=*]
    \item The \textit{Insert} strategy inserts the selected API sequence into a semantically compatible position within the current API sequence;
    \item The \textit{Replace} strategy replaces a randomly selected subsequence in the current API sequence with the selected API sequence, preserving type and semantic consistency;
    \item The \textit{Combine} strategy merges the selected API sequence with the API sequence of the current fuzz driver.
\end{itemize}
By applying these strategies, our method integrates high-potential API sequences associated with higher-potential APIs into the existing fuzz driver, thereby producing higher-quality drivers that are more likely to reach new execution paths or uncover potential vulnerabilities.

\section{Evaluation}

To evaluate the effectiveness of \toolname, we seek to answer the following research questions:

\begin{description}
\item [\textbf{RQ.1}] How much more code coverage does \toolname achieve compared to the baselines?

\item [\textbf{RQ.2}] Do the multidimensional API sequences contribute to the overall effectiveness of \toolname?

\item [\textbf{RQ.3}] Do the coverage-guided time scheduler and fuzz driver mutation strategy contribute to the overall effectiveness of \toolname?

\item [\textbf{RQ.4}] What is the impact of different LLMs on the performance of \toolname?

\item [\textbf{RQ.5}] Can \toolname find previously unknown bugs in the extensively tested benchmarks?
\end{description}

\subsection{Experimental Setup}

\begin{table*}[h]
\centering
\caption{Average code coverage achieved by our \toolname and the baselines \PromeFuzz, \CKGFuzzer and \LibErator.}
\label{tab:coverage}
\setlength{\tabcolsep}{2.5pt}
\renewcommand{\arraystretch}{1.1}
\begin{adjustbox}{width=\textwidth}
\small
\begin{tabular}{l c r r | r | rr | rr | rr}
\toprule
\multirow{2}{*}{\textbf{Library}} &
\multirow{2}{*}{\textbf{Version}} &
\multirow{2}{*}{\textbf{\#Branch}} &
\multirow{2}{*}{\textbf{\#API}} &
\multicolumn{7}{c}{\bfseries Code Coverage Comparison} \\
\cline{5-11}
& & & &
\textbf{MASFuzzer} &
\textbf{\PromeFuzz} & \textbf{Improv} &
\textbf{\CKGFuzzer} & \textbf{Improv} &
\textbf{\LibErator} & \textbf{Improv} \\
\midrule
c-ares & 42ddbc1 & 9166 & 151 & \textbf{5605.9} & 5285.1 & +6.07\% & 4743.4 & +18.18\% & 4947.8 & +13.30\%\\
cjson & 12c4bf1 & 1048 & 77 & \textbf{858.7} & 798.7 & +7.51\% & 824.4 & +4.17\% & 758.2 & +13.26\%\\
libpcap & f571971 & 7698 & 78 & \textbf{3360.9} & 2880.3 & +16.69\% & 2963.7 & +13.40\% & 3334.8 & +0.78\% \\
libtiff & 67c1cab & 12414 & 169 & 5118.3 & \textbf{5869.6} & -12.80\% & 4246.8 & +20.52\% & 3407.6 & +50.20\% \\
libvpx & 337f4bd & 33070 & 37 & \textbf{5279.8} & 4865.5 & +8.50\% & 3870.3 & +36.42\% & 2973.0 & +77.59\%\\
zlib & 5a82f71 & 2906 & 81 & \textbf{1909.5} & 1334.1 & +43.1\% & 1731.1 & +10.31\% & 1750.6 & +9.08\%\\
libssh2 & 694b9d9 & 8728 & 172 & \textbf{974.0} & 964.2 & +1.02\% & 955.7 & +1.92\% & 905.3 & +7.59\%\\
libplist & 20d5d57 & 3922 & 101 & \textbf{3087.4} & 2889.8 & +6.84\% & 2759.5 & +11.88\% & 2059.4 & +49.92\% \\
libass & 0f37982 & 5798 & 45 & \textbf{3724.1} & 3271.8 & +13.82\% & 3414.4 & +9.07\% & 3574.8 & +4.18\%\\
libzip & 542fc1c & 4406 & 106 & \textbf{2595.1} & 2476.2 & +4.80\% & 2500.4 & +3.79\% & 2271.9 & +14.23\% \\
libxlsxwriter & f6e8305 & 8806 & 300 & \textbf{5465.6} & 5443.9 & +0.40\% & 4106.5 & +33.10\% & 1935.7 & +182.36\%\\
stormlib & 49b619b & 10630 & 76 & \textbf{3527.7} & 3313.2 & +6.47\% & 2987.3 & +18.09\% & 2654.1 & +32.92\% \\
\midrule
\multicolumn{4}{c|}{\textbf{Average}} & - & - & +8.54\% & - & +15.07\% & - & +37.95\%\\
\bottomrule
\end{tabular}
\end{adjustbox}
\end{table*}

\textbf{Implementation.} We implemented \toolname with over 10K lines of Python code. For static analysis in \textit{metainfo} extraction, we employed Tree-sitter to parse the abstract syntax tree (AST). Regarding all LLMs, we selected the Deepseek-V3 as the default model due to its strong performance and cost efficiency. 
For fuzzing, we adopted LibFuzzer and performed instrumentation through multiple Sanitizers, including ASan and MSan, to detect runtime program errors.

\textbf{Comparison Baselines.} As baselines for comparison, we selected three state-of-the-art library fuzzers: \LibErator~\cite{toffalini2025liberating}, representing traditional (non-LLM-based) library fuzzers, \PromeFuzz~\cite{promefuzz-ccs25} and \CKGFuzzer~\cite{xu2025ckgfuzzer}, representing LLM-based library fuzzers. Among traditional library fuzzers, \LibErator is the most recent work and demonstrates the best coverage performance. All other traditional library fuzzers have previously been unfavorably compared to \LibErator. For LLM-based library fuzzers, \PromeFuzz and \CKGFuzzer are the most recent works, which significantly outperform both OSS-Fuzz-Gen and \PromptFuzz; therefore, we include \PromeFuzz and \CKGFuzzer in our evaluation. For \PromeFuzz, we adopt its recommended fuzzer, AFL++. For \CKGFuzzer, we limit the size of its constructed graph to avoid fuzz driver construction failures caused by exceeding the LLM context length. For \LibErator, both the generation time and the testing time use the recommended optimal configurations. For libraries where no optimal configuration is specified, we uniformly adopt a 6-hour generation and 18-hour testing setup.

\textbf{Benchmark Libraries.} We selected 12 widely used real-world C libraries for evaluation, including \texttt{c-ares}, \texttt{cjson}, \texttt{libpcap}, \texttt{libtiff}, \texttt{libvpx}, \texttt{zlib}, \texttt{libplist}, \texttt{libssh2}, \texttt{libass}, \texttt{libzip}, \texttt{libxlsxwriter} and \texttt{stormlib}. Among them, eight libraries are from the benchmark suites that are used for evaluating comparison fuzzers~\cite{sherman2025no, toffalini2025liberating, promefuzz-ccs25, lyu2024prompt, xu2025ckgfuzzer}. 
We also included four additional libraries from the OSS-Fuzz open-source project. We selected these libraries based on the project activity (with code or issue updates within the past month) and the community recognition (with more than 500 GitHub stars). 
For all selected libraries, we used the latest versions available on GitHub, shown in \autoref{tab:coverage}.

\textbf{Experimental Environment.} We evaluated \toolname within a Docker container running Ubuntu 20.04 on an Intel Xeon Gold 5218 @2.30GHz CPU with 128 GB of RAM. All libraries and fuzz drivers were compiled and executed in this environment, and the reported results are the average values obtained from ten independent runs. For comparison, we ran \CKGFuzzer and \PromeFuzz using the same Deepseek-V3 model. 

\textbf{Configuration Parameters.} In the API sequence construction stage, we set max length $L$ to 10. In the compilation repair stage, the maximum number of repair attempts is $3$. For time scheduling, the coverage threshold $\theta$ is set to $0.9$, the time factor bound $\beta$ to $0.2$, and the time coefficient base $B$ to $2.0$. In the fuzz driver mutation stage, the weighting coefficients $(\alpha, \beta, \gamma)$ are set to $(0.2, 0.05, 0.01)$, respectively. Additionally, both API-misuse crash repair and fuzz driver mutation are limited to one attempt per occurrence.

\subsection{RQ.1: Comparisons with Baselines}

We conducted 24-hour fuzzing campaigns on the selected target libraries using fuzz drivers generated by our tool \toolname, as well as baselines \PromeFuzz, \CKGFuzzer and \LibErator. Following the community recommendations~\cite{klees2018evaluating}, we report the average branch coverage for comparison. The overall coverage results are shown in \autoref{tab:coverage}. Here, we list the library version used (\textit{Version}), the total number of branches (\textit{\#Branch}) and APIs (\textit{\#APIs}) for each library, the branches covered by each tool, and the percentage improvement of our \toolname over the baselines (\textit{Improv}).

\textbf{Overall Results.} Compared to all baselines, \toolname covered a greater number of code branches. On average, \toolname covered 8.54\% more code than \PromeFuzz, reaching up to 16.69\% on the library \texttt{libpcap}. Similarly, \toolname achieves 15.07\% and 37.95\% higher coverage than \CKGFuzzer and \LibErator, respectively. It is worth noting that for more complex libraries with a larger number of branches, such as \texttt{libvpx}, and \texttt{libxlsxwriter}, the improvements of \toolname over baselines were more significant. This highlights the effectiveness of \toolname in exploring deep execution branches under complex scenarios. 

\begin{figure*}[h!]
    \centering
    \includegraphics[width=\textwidth]{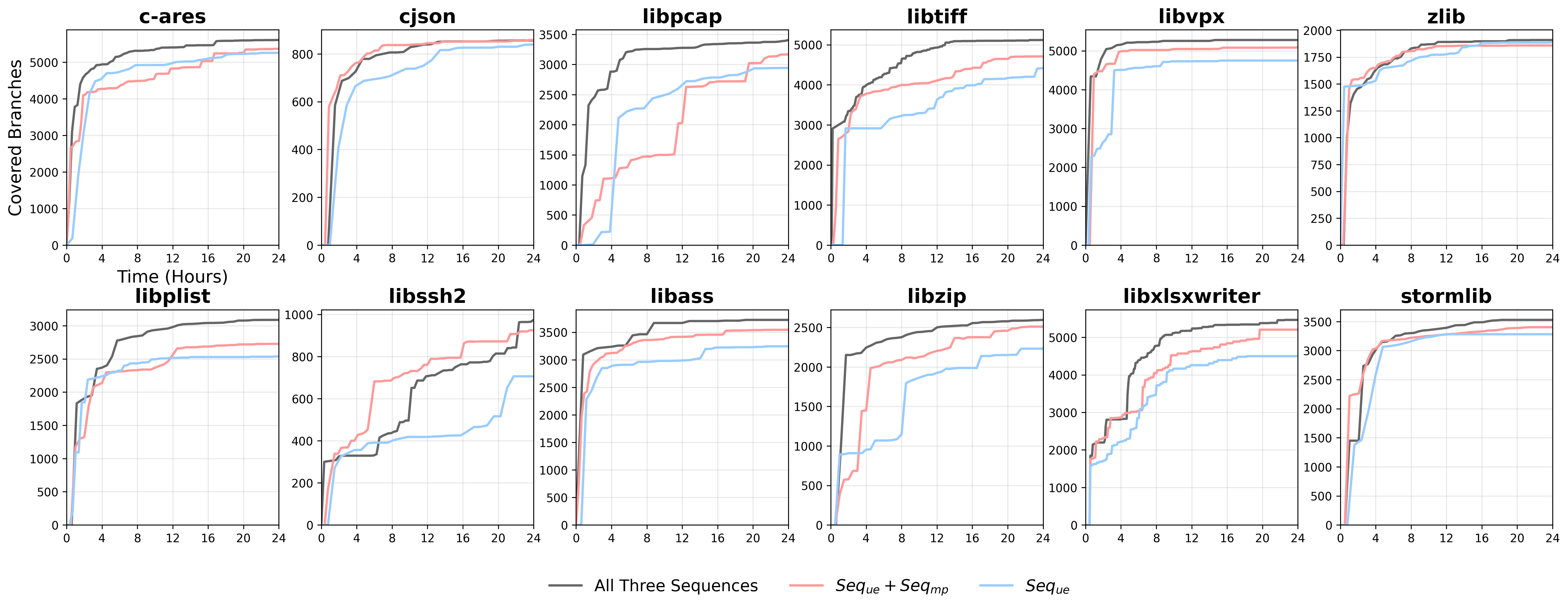}
    \caption{Ablation results of \toolname with different dimensions of API sequences. }
    \label{fig:ablation1}
\end{figure*}

\textbf{Comparing Results with \PromeFuzz.} \toolname outperforms \PromeFuzz in 11 out of the 12 target libraries.
Adopting a knowledge-driven paradigm, \PromeFuzz constructs a knowledge base incorporating metadata and documentation constraints, schedules highly correlated API sets based on multidimensional relevance (type, scope, and semantics), and generates drivers assisted by RAG. While it shares similarities with \toolname in leveraging consumer code and type matching to capture API relationships, a critical limitation remains: the extracted loose correlations cannot be directly mapped to effective execution sequences. Furthermore, lacking the driver-level time scheduling and mutation strategies inherent to \toolname, \PromeFuzz is unable to further exploit API relationships to optimize exploration after generation. Instead, it relies on the random scheduling of the underlying fuzzer, rendering the effectiveness of its RAG-generated fuzz drivers heavily dependent on the quality of retrieved documentation. Specifically, \PromeFuzz demonstrates strong competitiveness in libraries with extensive documentation resources. For instance, in \texttt{libtiff} and \texttt{libxlsxwriter}, which possess 110k+ and 100k+ characters of documentation respectively, \PromeFuzz achieves 12.80\% higher branch coverage than \toolname in the former, and trails by a negligible 0.4\% in the latter. Conversely, when applied to \texttt{zlib}, which contains only 8k characters of documentation of API comments, the performance of \PromeFuzz degrades significantly, resulting in a coverage gap of 43.1\% compared to \toolname.

\textbf{Comparing Results with \CKGFuzzer.} Across all target libraries, \toolname demonstrates consistently higher code coverage than \CKGFuzzer. Both are LLM-based fuzz-driver generation frameworks, but \CKGFuzzer relies on code knowledge graphs to extract API invocation relationships. This result confirms our observation that the correlation between call graphs and valid API sequences is relatively weak, and only relying on the API invocation relationship is not enough to capture sufficient API sequences. In addition, \CKGFuzzer relies solely on the LLM’s semantic understanding to construct fuzz drivers from the retrieved API combinations. Such weak constraints are prone to inducing API misuse. In contrast, \toolname effectively avoids this limitation by deeply modeling both the type constraints and semantic relationships among APIs. Moreover, code knowledge graphs often introduce excessive redundant information. In our evaluation across 12 target libraries, the prompts for 8 libraries exceeded the 64K context window limit due to the inclusion of large-scale call graphs. This results in context dilution, which further hinders the model’s ability to capture critical API invocation information. Collectively, these factors substantially limit \CKGFuzzer’s effectiveness in generating high-quality fuzz drivers.

\textbf{Comparing Results with \LibErator.} Across all target libraries, \toolname achieves markedly higher code coverage than \LibErator. Similar to \toolname, \LibErator leverages type matching to establish API dependencies, synthesizes candidate fuzz drivers by traversing API Flow Graphs (AFGs), and applies clustering-based algorithms for driver selection. However, due to its consumer-agnostic design, \LibErator lacks grounding in concrete usage patterns as ground truth and struggles to infer implicit, semantically driven logical dependencies among APIs. As a result, a portion of potentially reachable code paths remains unexplored, preventing \LibErator from effectively covering certain branches with high fuzzing potential.

In terms of LLM cost, the experiment for \toolname takes \$49.87, remaining more cost-effective than \CKGFuzzer (\$67.81).
\CKGFuzzer incurs substantial expenses due to the inclusion of extremely lengthy call graphs in its prompt context.
Conversely, while this exceeds \PromeFuzz (\$11.43), the additional cost is justified by our comprehensive approach: \toolname utilizes multidimensional API sequences to generate more fuzz drivers and employs LLM-based agents for driver mutation and crash analysis. 


\begin{table*}[h!]
\centering
\small
\renewcommand{\arraystretch}{1.1}
\setlength{\tabcolsep}{2.5pt}
\caption{Branch coverage of \toolname with and without the time scheduler and mutation strategy.}
\begin{tabularx}{\textwidth}{
l
|*{11}{>{\centering\arraybackslash}X<{\hsize=0.85\hsize}}
>{\centering\arraybackslash}X<{\hsize=2\hsize}
|>{\centering\arraybackslash}X<{\hsize=1.1\hsize}
}
\toprule
& \textbf{c-ares} & \textbf{cjson} & \textbf{libpcap} & \textbf{libtiff} & \textbf{libvpx} & \textbf{zlib} & \textbf{libssh2} & \textbf{libplist} & \textbf{libass} & \textbf{libzip} & \textbf{stormlib} & \textbf{libxlsxwriter} & \textbf{Avg}\\
\midrule
\textbf{\toolname} & \textbf{5605.9} & \textbf{858.7} & \textbf{3360.9} & \textbf{5118.3} & \textbf{5279.8} & \textbf{1909.5} & \textbf{974.0} & \textbf{3087.4} & \textbf{3724.1} & \textbf{2595.1} & \textbf{3527.7} & \textbf{5465.6} & \textbf{3458.9}\\
\textbf{w/o Time Scheduler} & 5023.2 & 801.4 & 2977.6 & 4838.0 & 4783.2 & 1853.2 & 923.4 & 2824.3 & 3282.2 & 2421.8 & 3116.6 & 4935.7 & 3148.4\\
\textbf{w/o Mutation} & 5414.2 & 823.9 & 3315.7 & 4993.0 & 4862.4 & 1800.4 & 951.8 & 2977.5 & 3457.1 & 2487.0 & 3339.8 & 5291.6 & 3309.5\\
\bottomrule
\end{tabularx}
\label{tab:ablation}
\end{table*}

\begin{table*}[h!]
\centering
\small
\renewcommand{\arraystretch}{1.1} 
\setlength{\tabcolsep}{2.5pt}
\caption{Branch coverage achieved by \toolname under different models: Deepseek-V3 and GPT-5.}
\begin{tabularx}{\textwidth}{
    l
    |*{11}{>{\centering\arraybackslash}X<{\hsize=0.85\hsize}}
    >{\centering\arraybackslash}X<{\hsize=2\hsize}
    |>{\centering\arraybackslash}X<{\hsize=1.1\hsize}
}
\toprule
\toolname-Model & \textbf{c-ares} & \textbf{cjson} & \textbf{libpcap} & \textbf{libtiff} & \textbf{libvpx} & \textbf{zlib} & \textbf{libssh2} & \textbf{libplist} & \textbf{libass} & \textbf{libzip} & \textbf{stormlib} & \textbf{libxlsxwriter}  & \textbf{Avg}\\
\midrule
Deepseek-V3 & \textbf{5605.9} & \textbf{858.7} & 3360.9 & 5118.3 & 5279.8 & 1909.5 & 974.0 & 3087.4 & 3724.1 & \textbf{2595.1} & \textbf{3527.7} & 5465.6 & 3458.9 \\
GPT-5       & 5520.8 & 840.7 & \textbf{3564.9} & \textbf{5134.4} & \textbf{5107.3} & \textbf{1949.9} & \textbf{983.6} & \textbf{3103.9} & \textbf{3881.2} & 2538.7 & 3461.3 & \textbf{5487.1} & \textbf{3464.5}\\
\bottomrule
\end{tabularx}
\label{tab:models}
\end{table*}

\subsection{RQ.2: Contribution of Different API Sequence Dimensions}

\toolname adopts three dimensions of API sequences to guide fuzz-driver generation from usage examples, mutation-propagation analysis, and semantic-aware analysis. To evaluate the contribution of each dimension, we conducted an ablation study. Specifically, we developed two additional variant tools: \toolname using only API sequences from usage example (\( Seq_{ue} \)), \toolname using both API sequences from usage examples and mutation-propagation analysis  (\( Seq_{ue} + Seq_{mp}\)), and compared them with \toolname using all API sequences (\textit{All Three Sequences}). The results are shown in \autoref{fig:ablation1}.

\textbf{Overall Results.} Across all target libraries, \toolname using all dimensions of API sequences achieves the highest coverage. The variant using API sequences from both usage examples and mutation-propagation analysis ranked second, with an average coverage $5.17\%$ less than \toolname. In contrast, the variant relying solely on \seqone performed the worst, with an average coverage $11.55\%$ lower. These results are consistent with our theoretical analysis. 

\textbf{Detailed Analysis.} Upon further analysis of these subjects, we have the following findings. For libraries that have a larger number of test cases and usage scenarios, such as \texttt{zlib}, using \seqone only can achieve good code coverage, which is unsurprising. 
It is interesting to find that the effectiveness of \seqtwo largely depends on the extent to which the APIs within a library share common data structures. For example, in the \texttt{cjson} library, 73 out of its 78 APIs depend on the \texttt{cjson} structure. This characteristic makes mutation-propagation analysis particularly effective, which can generate complex API sequences that cover nearly all usage scenarios. In contrast, in the \texttt{c-ares} library, most APIs pass complex structures through pointer parameters while returning only simple status or result codes. Of its 152 APIs, only 48 produce return values that can be matched with other APIs’ parameters. In such cases, generating API sequences through semantic-aware analysis becomes necessary.

\subsection{RQ.3: Contribution of Time Scheduler and Mutation Strategy}


In this section, we investigate the contribution of the time scheduler and mutation strategy to the effectiveness of \toolname. To this end, we construct two variants of \toolname by disabling the time scheduler and the mutation strategy, respectively, and conduct an additional case study. In the variant without the time scheduler, each fuzz driver is allocated an equal testing duration. In the variant without the mutation strategy, any fuzz driver that fails to discover new coverage branches is immediately terminated and is not subjected to further mutation iterations for generating new fuzz drivers. The experimental results are presented in \autoref{tab:ablation}.

\textbf{Overall Results.} \toolname with both the time scheduler and mutation strategy enabled achieves higher coverage across all 12 tested libraries. On average, it improves coverage by 9.08\% over the variant without the time scheduler and by 4.57\% over the variant without the mutation strategy. These results indicate that the combination of coverage-guided time scheduling and fuzz driver mutation not only transforms ineffective fuzz drivers into ones with testing potential, but also efficiently allocates more computational resources to promising fuzz drivers, thereby enabling exploration of deeper program branches.



\subsection{RQ.4: Impact of Different LLMs}

To evaluate the impact of different models on \toolname, we compared the latest released model GPT-5 with DeepSeek-V3---the default model used by \toolname. We measured the code coverage achieved by each model over a 24-hour fuzzing period, and the results are shown in \autoref{tab:models}. In addition, we analyzed the style differences in fuzz drivers generated by these models.

\textbf{Impact on Code Coverage.} \toolname powered by GPT-5 achieved higher coverage in 8 out of the 12 tested libraries. However, the advantage was marginal, with an average improvement of only 0.16\%. In the case of \texttt{libpcap}, where the difference was the largest, GPT-5 exceeded Deepseek-V3 by 6.10\%. These findings indicate that, within the context of \toolname, the overall performance of GPT-5 and Deepseek-V3 is largely comparable.

\textbf{Impact on Driver Style.} Although their final coverage results were comparable, the styles of the generated drivers differed considerably. GPT-5 tends to produce hierarchical and structured code that explicitly distinguishes multidimensional API sequence invocation paths, emphasizes temporal ordering and dependency management, and adopts a defensive programming paradigm with extensive error handling and fallback mechanisms. In contrast, DeepSeek-V3 follows a concise three-stage structure consisting of initialization, invocation, and cleanup, generating compact linear execution code that focuses on exploring a single API path in depth and relies on a fail-fast error handling strategy. These differences highlight the distinct optimization objectives of the two models in code generation: GPT-5 prioritizes robustness and breadth of coverage, while DeepSeek-V3 emphasizes simplicity and execution efficiency.


\subsection{RQ.5: Bug Detection Capability}
\label{subsec:rq4}

We further evaluate the capability of \toolname in detecting real-world bugs and distinguishing true bugs from API misuse. This section presents the discovered vulnerabilities, representative case studies, and an analysis of false positives.

\begin{table}[t]
\centering
\small
\caption{New bugs found by \toolname}
\label{tab:promptfuzz-bugs}
\setlength{\tabcolsep}{4pt}
\renewcommand{\arraystretch}{1}
\begin{tabular}{p{0.04\columnwidth}  
                p{0.18\columnwidth}  
                p{0.42\columnwidth}  
                p{0.21\columnwidth}} 
\toprule
\textbf{ID} & \textbf{Library} & \textbf{Bug Type} & \textbf{Status} \\
\midrule
1  & cjson         & Circular Reference        & Confirmed \\
2  & libtiff       & NULL Pointer Dereference  & Fixed \\
3  & libtiff       & NULL Pointer Dereference  & Reported \\
4  & libssh2       & NULL Pointer Dereference  & Fixed \\
5  & libssh2       & NULL Pointer Dereference  & Fixed$^{\dagger}$ \\
6  & libssh2       & Memory Leak               & Fixed$^{\dagger}$ \\
7  & libssh2       & Buffer Overflow           & Fixed$^{\dagger}$ \\
8  & libssh2       & Memory Leak               & Confirmed$^{\dagger}$ \\
9  & libplist      & Circular Reference        & Fixed \\
10 & libplist      & Use-After-Free            & Fixed \\
11 & libplist      & Buffer Overflow           & Reported \\
12 & libxlsxwriter & Buffer Overflow           & Fixed$^{\dagger}$ \\
13 & libxlsxwriter & Infinite Loop             & Fixed$^{\dagger}$ \\
14 & libxlsxwriter & Buffer Overflow           & Fixed$^{\dagger}$ \\
15 & libxlsxwriter & NULL Pointer Dereference  & Fixed$^{\dagger}$ \\
16 & stormlib      & Buffer Overflow           & Fixed$^{\dagger}$ \\
\bottomrule
\end{tabular}

\begin{minipage}{\columnwidth}
\footnotesize
$\dagger$ Indicates bugs that have been assigned CVE IDs.
\end{minipage}
\end{table}

\renewcommand{\arraystretch}{1}
\begin{table}[t]
\centering
\small
\caption{Number of detected bugs by \toolname}
\label{tab:library_stats}
\begin{tabularx}{\columnwidth}{l *{4}{>{\centering\arraybackslash}X}}
\toprule
\textbf{Library} & \textbf{UC} & \textbf{MC} & \textbf{RB} & \textbf{CB (FB)} \\
\midrule
c-ares           & 4  & 4  & 0 & 0 \\
cjson            & 4  & 3  & 1 & 1 (0) \\
libpcap          & 2  & 2  & 0 & 0 \\
libtiff          & 5  & 3  & 2 & 1 (1) \\
libvpx           & 3  & 3  & 0 & 0 \\
zlib             & 1  & 1  & 0 & 0 \\
libssh2          & 8  & 3  & 5 & 5 (4) \\
libplist         & 5  & 2  & 3 & 2 (2) \\
libass           & 2  & 2  & 0 & 0 \\
libzip           & 1  & 1  & 0 & 0 \\
libxlsxwriter    & 6  & 2  & 4 & 4 (4) \\
stormlib         & 2  & 1  & 1 & 1 (1) \\\hline
Total            & 43 & 27 & 16& 14 (12) \\
\bottomrule
\end{tabularx}
\textbf{UC}: Unique Crash. 
\textbf{MC}: API Misuse Crash. 
\textbf{RB}: Reported Bugs. 
\textbf{CB}: Confirmed Bugs. 
\textbf{FB}: Fixed Bugs.
\end{table}

\textbf{Real-world Bug Detection.} \autoref{tab:promptfuzz-bugs} summarizes 16 previously unknown bugs discovered by \toolname, 14 of which have already been confirmed and fixed. These bugs span multiple categories, such as null pointer dereference, memory leak, and buffer overflow. For example, in the \verb|libssh2| library, \toolname identified a memory leak in the \texttt{libssh2\_knownhost\_check} function caused by missing object deallocation in special cases; in the \texttt{libtiff} library, we detected that the \texttt{TIFFReadCustomDirectory} function lacked a null pointer check, which resulted in a crash when subsequently invoking \texttt{strlen}. Moreover, the minimal triggering API sequence lengths of these bugs range from one to eleven, demonstrating that \toolname is capable of uncovering diverse types of bugs across multiple levels of complexity. 

\begin{figure}[h!]
    \centering
    \begin{minted}[
        fontsize=\small,
        breaklines,
        linenos,
        numbersep=4pt,
        xleftmargin=2pt
    ]{c}
static int hostline(LIBSSH2_KNOWNHOSTS *hosts, const char *host, size_t hostlen, const char *key, size_t keylen){
    while((*key == ' ') || (*key == '\t')) {
        key++;
        keylen--;
    }
}
    \end{minted}
    \caption{Vulnerable code of a CVE-assigned bug in \texttt{libssh2}.}
    \label{fig:case1}
    
\end{figure}

\textbf{Case Study.} \autoref{fig:case1} illustrates the buffer-overflow vulnerability detected by \toolname in the \texttt{libssh2} library. \texttt{libssh2\_knownhost\_\allowbreak readline} is responsible for parsing individual entries in the SSH client \texttt{known\allowbreak\_hosts} file, which calls the \verb|hostline| function to read a line of text and extract structured information, skipping over empty fields in the process. However, due to the lack of a proper check on \texttt{keylen}, when consecutive empty fields occur, the key pointer can exceed the valid data range, leading to an out-of-bounds memory access and triggering a buffer overflow.

\textbf{API Misuse VS Actual Bug.} \autoref{tab:library_stats} presents the number of unique crashes detected by \toolname, including those caused by API misuse and actual bugs. In total, \toolname identified 43 unique crashes, of which 27 were API misuses. A thorough analysis of these API misuses revealed that 17 instances involved violations of API documentation, such as passing null pointers, mismatched buffer sizes, and inconsistent parameter types. It is worth noting that due to the lack of built-in input validation in \verb|C| library, the boundary between API misuse and actual bugs is often blurred. For example, in cases of crashes caused by passing null pointers, the developers of \texttt{libtiff} consider adding assertions to prevent such issues valuable, while the developers of \texttt{libass} believe that the validity of such inputs should be the responsibility of the developers. Given the absence of prior knowledge, these situations cannot be fully determined by the fuzzing framework alone. It is therefore necessary to consult the project documentation and engage in further research and discussions with the library developers to gain a deeper understanding.




\section{Discussions and Future Directions}

\textbf{Data Contamination.} With the recent advances in LLM-based library fuzzing, widely tested libraries such as \texttt{cjson} and \texttt{c-ares} are frequently included in fuzzing benchmarks, which may already have a large number of fuzz drivers. When these fuzz drivers appear in the training corpus of LLMs, they may introduce data contamination: the observed coverage advantage might not stem from the novel design of \toolname, but from the model memorization on high-coverage fuzz drivers encountered during training. To mitigate this issue, we employed the same model and parameters when comparing \toolname with another LLM-based tool, minimizing discrepancies introduced by potential contamination. In comparison with traditional tools, however, this limitation is largely unavoidable. Furthermore, among the ten tested libraries, only eight had appeared in previous benchmarks; we intentionally included four additional libraries that had not been used before, in order to further reduce the risk of contamination.

\textbf{Construction of API Sequences with Complex Branches and Loops.} During the construction of multidimensional API sequences, \toolname only produces straight-line API sequences, without any branches or loops. However, when generating fuzer drivers, LLMs can introduce control-flow structures based on these straight-line API sequences. While ignoring such loops and branches in the initial API sequences can lead to incomplete coverage of complex scenarios, this limitation is shared by nearly all existing techniques of automated fuzz-driver generation. 

\textbf{Support for Other Programming Languages.} \toolname currently supports testing libraries written in \verb|C|. However, in \toolname, aside from the initial static analysis phase for parsing \textit{metainfo}, the remaining workflow is largely language-agnostic. Its remaining analysis leverages the static analysis tools of Tree-sitter, which support multiple programming languages. Therefore, it is relatively straightforward to extend \toolname to support other languages. In future work, we plan to support commonly used library programming languages such as \verb|C++| and \verb|Rust|.

\section{Related Work}
\textbf{Automated Fuzz-Driver Generation.} Traditional approaches primarily rely on static or dynamic analysis to infer API usage patterns from source code \cite{yan2025sok,zhang2025rumono}. A significant focus has been on correctly sequencing API calls. For instance, GraphFuzz uses lifecycle-aware data-flow graphs to model API dependencies \cite{green2022graphfuzz}, while RULF traverses API dependency graphs specifically to fuzz Rust libraries \cite{jiang2021rulf}. These methods are effective at capturing explicit data-flow constraints, where the output of one API directly serves as the input to another. Other techniques have focused on addressing different aspects of the problem. \Hopper~\cite{chen2023hopper} introduces an interpretative fuzzing framework that learns API constraints without requiring prior knowledge. More recent works refine the generation and mutation process; for example, OGHARN~\cite{sherman2025no} composes and filters drivers in an oracle-guided manner, libErator~\cite{toffalini2025liberating} aims to balance testing efficiency and depth, and RIMFuzz~\cite{wang2025rimfuzz} dynamically adjusts mutation strategies based on real-time coverage feedback. However, these approaches remain fundamentally constrained by information obtainable through code analysis alone, and often struggle to capture implicit semantic dependencies, idiomatic usage patterns, and hidden preconditions—such as required initialization sequences—that are critical for exercising complex library behaviors.

\textbf{LLM-based Fuzz Driver Generation.} The remarkable code comprehension and generation capabilities of Large Language Models (LLMs) have recently been applied to fuzz-driver generation. \PromptFuzz~\cite{lyu2024prompt} pioneered an iterative, coverage-guided strategy, using fuzzing feedback to refine LLM prompts and progressively generate better drivers, but it lacks an explicit mechanism for capturing relationships among APIs. To address this context deficit, \CKGFuzzer~\cite{xu2025ckgfuzzer} enhances LLM performance with Retrieval-Augmented Generation (RAG), querying a Code Knowledge Graph to provide more library-specific information. While this incorporates more structural data, the caller-callee relationships captured in a knowledge graph do not necessarily represent a valid temporal sequence of API invocations. Similarly, \PromeFuzz uses a knowledge-driven approach with document-RAG to extract API semantics and select candidate APIs, but it still produces an unordered set without explicit structural constraints and lacks driver scheduling or mutation mechanisms for further optimization. \toolname directly addresses these limitations by providing the LLM with explicit and rich contextual information through its multidimensional API sequences. 

\section{Conclusion}

In this paper, we propose a method for high-quality fuzz-driver generation through multidimensional API sequences construction. This approach extracts API usage patterns from the codebase and leverages both API mutation-propagation-ground and semantic-aware mining to build effective API sequences during fuzz-driver generation. Based on this design, we develop \toolname, a fuzzing framework that integrates a dynamic fuzz driver time scheduler with a coverage-guided fuzz driver mutation strategy. We evaluated \toolname on 12 real-world libraries, and the results demonstrate that it outperforms existing tools in terms of code coverage while successfully detecting 16 previously unknown bugs in the extensively tested libraries. Overall, \toolname is capable of generating higher-quality fuzz drivers and provides a general and efficient solution for library fuzzing.

\section*{Data Availability}

The data and source code used in this paper are available at https://anonymous.4open.science/r/ano-3545/.

\bibliographystyle{IEEEtran}
\bibliography{main}

\begin{thebibliography}{10}
\providecommand{\url}[1]{#1}
\csname url@samestyle\endcsname
\providecommand{\newblock}{\relax}
\providecommand{\bibinfo}[2]{#2}
\providecommand{\BIBentrySTDinterwordspacing}{\spaceskip=0pt\relax}
\providecommand{\BIBentryALTinterwordstretchfactor}{4}
\providecommand{\BIBentryALTinterwordspacing}{\spaceskip=\fontdimen2\font plus
\BIBentryALTinterwordstretchfactor\fontdimen3\font minus \fontdimen4\font\relax}
\providecommand{\BIBforeignlanguage}[2]{{%
\expandafter\ifx\csname l@#1\endcsname\relax
\typeout{** WARNING: IEEEtran.bst: No hyphenation pattern has been}%
\typeout{** loaded for the language `#1'. Using the pattern for}%
\typeout{** the default language instead.}%
\else
\language=\csname l@#1\endcsname
\fi
#2}}
\providecommand{\BIBdecl}{\relax}
\BIBdecl

\bibitem{fioraldi2020aflpp}
\BIBentryALTinterwordspacing
A.~Fioraldi, D.~Maier, H.~Eißfeldt, and M.~Heuse, ``Afl++: Combining incremental steps of fuzzing research,'' in \emph{14th USENIX Workshop on Offensive Technologies (WOOT 20)}.\hskip 1em plus 0.5em minus 0.4em\relax USENIX Association, 2020. [Online]. Available: \url{https://www.usenix.org/conference/woot20/presentation/fioraldi}
\BIBentrySTDinterwordspacing

\bibitem{serebryany2016continuous}
K.~Serebryany, ``Continuous fuzzing with libfuzzer and addresssanitizer,'' in \emph{IEEE Cybersecurity Development Conference (SecDev)}, 2016.

\bibitem{bohme2020boosting}
M.~B{\"o}hme, V.~J.~M. Man{\`e}s, and S.~K. Cha, ``Boosting fuzzer efficiency: An information theoretic perspective,'' in \emph{ACM Joint Meeting on Foundations of Software Engineering (FSE)}.\hskip 1em plus 0.5em minus 0.4em\relax ACM, 2020.

\bibitem{manes2018art}
V.~J. Manes, H.~Han, C.~Han, S.~K. Cha, M.~Egele, E.~J. Schwartz, and M.~Woo, ``The art, science, and engineering of fuzzing: A survey,'' \emph{arXiv preprint arXiv:1812.00140}, 2018.

\bibitem{chafjiri2024vulnerability}
S.~B. Chafjiri, P.~Legg, J.~Hong, and M.-A. Tsompanas, ``Vulnerability detection through machine learning-based fuzzing: A systematic review,'' \emph{Computers \& Security}, vol. 143, p. 103903, 2024.

\bibitem{ispoglou2020fuzzgen}
K.~Ispoglou, D.~Austin, V.~Mohan, and M.~Payer, ``$\{$FuzzGen$\}$: Automatic fuzzer generation,'' in \emph{29th USENIX Security Symposium (USENIX Security 20)}, 2020, pp. 2271--2287.

\bibitem{liu2024afgen}
Y.~Liu, Y.~Wang, X.~Jia, Z.~Zhang, and P.~Su, ``Afgen: Whole-function fuzzing for applications and libraries,'' in \emph{2024 IEEE Symposium on Security and Privacy (SP)}.\hskip 1em plus 0.5em minus 0.4em\relax IEEE, 2024, pp. 1901--1919.

\bibitem{serebryany2017oss}
K.~Serebryany, ``{OSS-Fuzz}: Google's continuous fuzzing service for open source software,'' in \emph{USENIX Security Symposium}, 2017.

\bibitem{liu2025llm4tdg}
J.~Liu, R.~Liang, X.~Zhu, Y.~Zhang, Y.~Liu, and Q.~Liu, ``Llm4tdg: test-driven generation of large language models based on enhanced constraint reasoning,'' \emph{Cybersecurity}, vol.~8, no.~1, p.~32, 2025.

\bibitem{jeong2023utopia}
B.~Jeong, J.~Jang, H.~Yi, J.~Moon, J.~Kim, I.~Jeon, T.~Kim, W.~Shim, and Y.~H. Hwang, ``Utopia: Automatic generation of fuzz driver using unit tests,'' in \emph{2023 IEEE Symposium on Security and Privacy (SP)}.\hskip 1em plus 0.5em minus 0.4em\relax IEEE, 2023, pp. 2676--2692.

\bibitem{babic2019fudge}
D.~Babi{\'c}, S.~Bucur, Y.~Chen, F.~Ivan{\v{c}}i{\'c}, T.~King, M.~Kusano, C.~Lemieux, L.~Szekeres, and W.~Wang, ``Fudge: fuzz driver generation at scale,'' in \emph{Proceedings of the 2019 27th ACM Joint Meeting on European Software Engineering Conference and Symposium on the Foundations of Software Engineering}, 2019, pp. 975--985.

\bibitem{green2022graphfuzz}
H.~Green and T.~Avgerinos, ``Graphfuzz: Library api fuzzing with lifetime-aware dataflow graphs,'' in \emph{Proceedings of the 44th International Conference on Software Engineering}, 2022, pp. 1070--1081.

\bibitem{mallissery2023demystify}
S.~Mallissery and Y.-S. Wu, ``Demystify the fuzzing methods: A comprehensive survey,'' \emph{ACM Computing Surveys}, vol.~56, no.~3, pp. 71:1--71:38, 2023.

\bibitem{Liu_OSS-Fuzz-Gen_Automated_Fuzz_2024}
\BIBentryALTinterwordspacing
D.~Liu, O.~Chang, J.~metzman, M.~Sablotny, and M.~Maruseac, ``{OSS-Fuzz-Gen: Automated Fuzz Target Generation},'' May 2024. [Online]. Available: \url{https://github.com/google/oss-fuzz-gen}
\BIBentrySTDinterwordspacing

\bibitem{lyu2024prompt}
Y.~Lyu, Y.~Xie, P.~Chen, and H.~Chen, ``Prompt fuzzing for fuzz driver generation,'' in \emph{Proceedings of the 2024 ACM SIGSAC Conference on Computer and Communications Security}, 2024, pp. 3793--3807.

\bibitem{xu2025ckgfuzzer}
H.~Xu, W.~Ma, T.~Zhou, Y.~Zhao, K.~Chen, Q.~Hu, Y.~Liu, and H.~Wang, ``Ckgfuzzer: Llm-based fuzz driver generation enhanced by code knowledge graph,'' in \emph{2025 IEEE/ACM 47th International Conference on Software Engineering: Companion Proceedings (ICSE-Companion)}.\hskip 1em plus 0.5em minus 0.4em\relax IEEE, 2025, pp. 243--254.

\bibitem{promefuzz-ccs25}
\BIBentryALTinterwordspacing
Y.~Liu, J.~Deng, X.~Jia, Y.~Wang, M.~Wang, L.~Huang, T.~Wei, and P.~Su, ``Promefuzz: A knowledge-driven approach to fuzzing harness generation with large language models,'' in \emph{Proceedings of the 2025 ACM SIGSAC Conference on Computer and Communications Security}, ser. CCS '25.\hskip 1em plus 0.5em minus 0.4em\relax New York, NY, USA: Association for Computing Machinery, 2025, p. 1559–1573. [Online]. Available: \url{https://doi.org/10.1145/3719027.3765222}
\BIBentrySTDinterwordspacing

\bibitem{toffalini2025liberating}
F.~Toffalini, N.~Badoux, Z.~Tsinadze, and M.~Payer, ``Liberating libraries through automated fuzz driver generation: Striking a balance without consumer code,'' \emph{Proceedings of the ACM on Software Engineering}, vol.~2, no. FSE, pp. 2123--2145, 2025.

\bibitem{bohme2020fuzzing}
M.~B{\"o}hme, C.~Cadar, and A.~Roychoudhury, ``Fuzzing: Challenges and reflections,'' \emph{IEEE Software}, vol.~38, no.~3, pp. 79--86, 2020.

\bibitem{zhao2024systematic}
X.~Zhao, H.~Qu, J.~Xu, X.~Li, W.~Lv, and G.-G. Wang, ``A systematic review of fuzzing,'' \emph{Soft Computing}, vol.~28, no.~6, pp. 5493--5522, 2024.

\bibitem{fioraldi2023dissecting}
A.~Fioraldi, A.~Mantovani, D.~Maier, and D.~Balzarotti, ``Dissecting american fuzzy lop: a fuzzbench evaluation,'' \emph{ACM transactions on software engineering and methodology}, vol.~32, no.~2, pp. 1--26, 2023.

\bibitem{pham2019smart}
V.-T. Pham, M.~B{\"o}hme, A.~E. Santosa, A.~R. C{\u{a}}ciulescu, and A.~Roychoudhury, ``Smart greybox fuzzing,'' \emph{IEEE Transactions on Software Engineering}, vol.~47, no.~9, pp. 1980--1997, 2019.

\bibitem{wang2019superion}
J.~Wang, B.~Chen, L.~Wei, and Y.~Liu, ``Superion: Grammar-aware greybox fuzzing,'' in \emph{2019 IEEE/ACM 41st International Conference on Software Engineering (ICSE)}.\hskip 1em plus 0.5em minus 0.4em\relax IEEE, 2019, pp. 724--735.

\bibitem{libfuzzer}
{LLVM Project}, ``Libfuzzer --- a library for coverage-guided fuzz testing,'' \url{https://llvm.org/docs/LibFuzzer.html}, accessed: 2026-03-29.

\bibitem{achiam2023gpt}
J.~Achiam, S.~Adler, S.~Agarwal, L.~Ahmad, I.~Akkaya, F.~L. Aleman, D.~Almeida, J.~Altenschmidt, S.~Altman, S.~Anadkat \emph{et~al.}, ``Gpt-4 technical report,'' \emph{arXiv preprint arXiv:2303.08774}, 2023.

\bibitem{roziere2023code}
B.~Roziere, J.~Gehring, F.~Gloeckle, S.~Sootla, I.~Gat, X.~E. Tan, Y.~Adi, J.~Liu, R.~Sauvestre, T.~Remez \emph{et~al.}, ``Code llama: Open foundation models for code,'' \emph{arXiv preprint arXiv:2308.12950}, 2023.

\bibitem{chen2021evaluating}
M.~Chen, J.~Tworek, H.~Jun, Q.~Yuan, H.~P. D.~O. Pinto, J.~Kaplan, H.~Edwards, Y.~Burda, N.~Joseph, G.~Brockman \emph{et~al.}, ``Evaluating large language models trained on code,'' \emph{arXiv preprint arXiv:2107.03374}, 2021.

\bibitem{cheng2025towards}
Y.~Cheng, H.~J. Kang, L.~K. Shar, C.~Dong, Z.~Shi, S.~Lv, and L.~Sun, ``Towards reliable llm-driven fuzz testing: Vision and road ahead,'' \emph{arXiv preprint arXiv:2503.00795}, 2025.

\bibitem{wang2023review}
J.~Wang and Y.~Chen, ``A review on code generation with llms: Application and evaluation,'' in \emph{2023 IEEE International Conference on Medical Artificial Intelligence (MedAI)}.\hskip 1em plus 0.5em minus 0.4em\relax IEEE, 2023, pp. 284--289.

\bibitem{siddiq2024using}
M.~L. Siddiq, J.~C. Da~Silva~Santos, R.~H. Tanvir, N.~Ulfat, F.~Al~Rifat, and V.~Carvalho~Lopes, ``Using large language models to generate junit tests: An empirical study,'' in \emph{Proceedings of the 28th international conference on evaluation and assessment in software engineering}, 2024, pp. 313--322.

\bibitem{schafer2024empirical}
M.~Sch{\"a}fer, S.~Nadi, A.~Eghbali, and F.~Tip, ``An empirical evaluation of using large language models for automated unit test generation,'' \emph{IEEE Transactions on Software Engineering}, vol.~50, no.~1, pp. 85--105, 2024.

\bibitem{liu2025can}
J.~Liu, S.~Lee, E.~Losiouk, and M.~B{\"o}hme, ``Can llm generate regression tests for software commits?'' \emph{arXiv preprint arXiv:2501.11086}, 2025.

\bibitem{xia2023keep}
C.~S. Xia and L.~Zhang, ``Keep the conversation going: Fixing 162 out of 337 bugs for \$0.42 each using chatgpt,'' \emph{arXiv preprint arXiv:2304.00385}, 2023.

\bibitem{xia2024fuzz4all}
C.~S. Xia, M.~Paltenghi, J.~L. Tian, M.~Pradel, and L.~Zhang, ``Fuzz4all: Universal fuzzing with large language models,'' in \emph{Proceedings of the 46th IEEE/ACM International Conference on Software Engineering}, 2024, pp. 126:1--126:13.

\bibitem{meng2024large}
R.~Meng, M.~Mirchev, M.~B{\"o}hme, and A.~Roychoudhury, ``Large language model guided protocol fuzzing,'' in \emph{Proceedings of the 31st Annual Network and Distributed System Security Symposium (NDSS)}, vol. 2024, 2024.

\bibitem{hu2023augmenting}
J.~Hu, Q.~Zhang, and H.~Yin, ``Augmenting greybox fuzzing with generative ai,'' \emph{arXiv preprint arXiv:2306.06782}, 2023.

\bibitem{eom2024fuzzing}
J.~Eom, S.~Jeong, and T.~Kwon, ``Fuzzing javascript interpreters with coverage-guided reinforcement learning for llm-based mutation,'' in \emph{Proceedings of the 33rd ACM SIGSOFT International Symposium on Software Testing and Analysis}, 2024, pp. 1656--1668.

\bibitem{deng2023large}
Y.~Deng, C.~S. Xia, H.~Peng, C.~Yang, and L.~Zhang, ``Large language models are zero-shot fuzzers: Fuzzing deep-learning libraries via large language models,'' in \emph{Proceedings of the 32nd ACM SIGSOFT international symposium on software testing and analysis}, 2023, pp. 423--435.

\bibitem{chen2023hopper}
P.~Chen, Y.~Xie, Y.~Lyu, Y.~Wang, and H.~Chen, ``Hopper: Interpretative fuzzing for libraries,'' in \emph{Proceedings of the 2023 ACM SIGSAC Conference on Computer and Communications Security}, 2023, pp. 1600--1614.

\bibitem{zhu2022fuzzing}
X.~Zhu, S.~Wen, S.~Camtepe, and Y.~Xiang, ``Fuzzing: A survey for roadmap,'' \emph{ACM Computing Surveys (CSUR)}, vol.~54, no. 11s, pp. 1--36, 2022.

\bibitem{bohme2016coverage}
M.~B{\"o}hme, V.-T. Pham, and A.~Roychoudhury, ``Coverage-based greybox fuzzing as markov chain,'' in \emph{Proceedings of the 2016 ACM SIGSAC Conference on Computer and Communications Security}, 2016, pp. 1032--1043.

\bibitem{sherman2025no}
\BIBentryALTinterwordspacing
G.~Sherman and S.~Nagy, ``No harness, no problem: Oracle-guided harnessing for auto-generating c api fuzzing harnesses,'' in \emph{2025 IEEE/ACM 47th International Conference on Software Engineering (ICSE)}.\hskip 1em plus 0.5em minus 0.4em\relax IEEE Computer Society, 2025, pp. 165--177. [Online]. Available: \url{https://doi.org/10.1109/ICSE55347.2025.00239}
\BIBentrySTDinterwordspacing

\bibitem{klees2018evaluating}
G.~Klees, A.~Ruef, B.~Cooper, S.~Wei, and M.~Hicks, ``Evaluating fuzz testing,'' in \emph{Proceedings of the 2018 ACM SIGSAC conference on computer and communications security}, 2018, pp. 2123--2138.

\bibitem{yan2025sok}
Q.~Yan, M.~Huang, H.~Cao, and S.~Lu, ``Sok: From systematization to best practices in fuzz driver generation,'' in \emph{Australasian Conference on Information Security and Privacy}.\hskip 1em plus 0.5em minus 0.4em\relax Springer, 2025, pp. 348--368.

\bibitem{zhang2025rumono}
Y.~Zhang, J.~Wu, and H.~Xu, ``Rumono: Fuzz driver synthesis for rust generic apis,'' \emph{ACM Transactions on Software Engineering and Methodology}, vol.~34, no.~6, pp. 169:1--169:28, 2025.

\bibitem{jiang2021rulf}
J.~Jiang, H.~Xu, and Y.~Zhou, ``Rulf: Rust library fuzzing via api dependency graph traversal,'' in \emph{2021 36th IEEE/ACM International Conference on Automated Software Engineering (ASE)}.\hskip 1em plus 0.5em minus 0.4em\relax IEEE, 2021, pp. 581--592.

\bibitem{wang2025rimfuzz}
X.~Wang and L.~Zhao, ``Rimfuzz: real-time impact-aware mutation for library api fuzzing,'' \emph{Journal of King Saud University Computer and Information Sciences}, vol.~37, no.~4, p.~52, 2025.

\end{thebibliography}

\end{document}